\documentstyle[graphicx]{article}

\newcommand{\degree}{\mbox{$^{\circ}$}}

\begin{document}

\Large
{\bf Investigation of long-period variables in the Catalina Southern catalog: new carbon stars and false objects}\\

\large

N. Mauron$^1$, K.S. Gigoyan$^2$, K.K. Gigoyan$^3$, L.P.A. Maurin$^4$, T.R.~Kendall$^5$\\

{\small$^1$\,University of Montpellier, LUPM UMR 5299 (UM/CNRS) CC72, 34095 Montpellier, France, nicolas.mauron@umontpellier.fr

 $^2$\,NAS RA V.A. Ambartsumian Byurakan Astrophysical Observatory (BAO), Armenia, kgigoyan@bao.sci.am

$^3$\,Yerevan State University, Armenia

$^4$\,Observatoire des Ifs, 6 impasse des Ifs, 84000 Avignon, France

$^5$\,Northampton NN1 4RG, United Kingdom}\\

Accepted for publication in the Journal: {\it Astrophysics}  (Springer www.springer.com)

(Jan 28, 2019)

\large

\section{Abstract ({\it see the last page for abstract in russian)} }

As part of our ongoing study of the evolved giants in the galactic halo, we consider the  sample of 1286 long-period variables (LPVs) in the southern hemisphere provided by the Catalina Sky Survey experiment. These LPVs have periods P $>$ 80 days and amplitudes $>$  0.2 mag. First, by using the Hamburg-ESO spectral survey, we aim to determine the spectral type as either M-type or C-type for objects located in the imprint of this survey, $|b| > 30$\degree. Of 135 LPVs obeying this selection, we classified 93, and found only 2 new carbon stars. Secondly, we consider faint LPVs. We discovered that many lie at $\sim$ 1 arcmin from a bright Mira catalogued in the General Catalog of Variable Stars, with identical period. We study these suspicious cases in detail, and conclude that, for as many as 56 faint Catalina LPVs, their variability is due to contamination by light from the bright, neighbouring GCVS Mira: an instrumental artefact. We conclude that when dealing with distant, faint Miras in the Catalina catalog, researchers should pay attention to the polluting effects of neighbouring bright and variable objects.

\section{Introduction}

The galactic halo is one of the major components of our Galaxy. Although considerable attention has been given to its component stellar populations and their properties, the population of asymptotic giant (AGB) stars located in the halo deserves more attention. One method to find AGB stars is to search for carbon (C) stars.  Huxor and Grebel (2015, hereafter HG) [1]  synthesized our knowledge on the subject and list $\sim$ 200 C stars in the halo. The origin of the majority of them is the Sagittarius dwarf galaxy, which is tidally disrupted in its orbital path around the Milky Way. Another method, adopted in this work, is to consider samples of variable stars for which light periodicity is obvious, and long periods ($>$ 100 days) are measured. In this work, we consider the recent Catalina catalog (CSS) of southern periodic variables (Drake et al. 2017) [2].

In order to better know the sample of the 1286 LPVs presented in the Catalina survey, we performed two distinct investigations: firstly, examination of spectra of a small subsample of these objects, located at high galactic latitude ($|b| > 30$\degree). For these objects, a spectrum can often be found on the plates of the Hamburg-ESO objective prism survey (Engels et al. 2001 [3]; Christlieb et al. 2004 [4]). This investigation is carried out to determine if the list of HG is complete, or if it is still possible to find numerous new carbon stars.  The second investigation aimed to focus on faint stars of the Drake et al. LPV sample. Because some of them are very faint ($V \sim 17$), and assuming that they are Mira or SRa variables, their distances can be considerable, typically 100~kpc or more. Our approach and results are described below in more detail.

\section{Classification of Catalina southern LPVs lying at  $|b| > 30$\degree}

Before spectral classification is described, some properties of the Southern Catalina catalog (Drake et al. 2017)[2] are worth pointing out. The catalog of 1286 LPVs  represents 3.4 percent of the total of 37745 variables of all types in the complete database. It is interesting to note that in the release of the Catalina DR1 {\it northern} catalog (Drake et al. 2014)[5], there were 512 catalogued LPVs. This  indicates that either the southern experiment is more sensitive, or that the galactic plane has been covered to a greater extent than in the north, or both. 

Figure 1 is a map showing the distribution of the 1286 southern LPVs, shown as small circles. All have $\delta < -20$\degree.
Many Catalina LPVs are closer to the Milky Way plane than $|b| = 20$\degree. Thus, this sample is dominated by galactic disc stars, with a small fraction being actually at high galactic latitude. A large number of stars are detected in the Large Magellanic Cloud, and also in the Fornax and Sculptor dwarf galaxies. This demonstrates that this sample reaches distances as great as 140~kpc (Fornax).

Among the 1286 LPVs, many are relatively bright. For example, for a Catalina magnitude $V_{\rm CSS} < 13$, there are 579 stars. But there are also many faint stars: 315 have $V_{\rm CSS} > 15$, with 67 as faint as $V_{\rm CSS} = $ 17.0 to 19.0. With the assumption that these stars are Miras, or periodic semi-regulars, their absolute magnitude is around $M_v \approx -3$ (Drake et al. 2014)[5], and the distances of these 67 fainter stars would be of the order of 100~kpc or greater.
 
\subsection{Spectral classification}

In order to determine spectral classification, we use the Hamburg-ESO (HES) plate survey (on-line on the Hambourg Observatory web site). 
These plates are deep since the survey  reaches $B_J = 18.0$. However, this survey is restricted to high galactic latitude ($|b| > 30$\degree). In addition, HES spectra are centered in the blue region from $\sim$~3200 \AA\ to 5300~\AA, and this may induce a bias against the reddest objects. When examining HES spectra, the  M-type classification is based on the presence of TiO bands at 5167, 4954, 4854, 4762, and 4584 \AA. In contrast,  C-type classification is based on C$_2$ bands at 5165 and 4737 \AA. Although the linear resolution is 15\AA\ at H$\gamma$, the separation C or M-type is quite clear when the star is not too faint. In addition to the HES survey, we use information from the CDS database, and from the Sloan survey spectral database (for Data Release 
12 at https://www.sdss.org/dr12/spectro/). 

Figure 2 illustrates the HES spectra for stars KG-21 (a C-type star) and KG-40 (an M-type star).
In the field covered by the HES survey, there are 135 Catalina LPVs. Of these 135 objects, our examination on the plates yielded 93 objects. The catalog of these objects, named KG-01 to KG-93, will be made available at CDS. 

\subsection{Carbon stars}

Although most are M-type, five C-type stars were found. They cannot be dwarf carbon stars since dwarf stars are not large amplitude periodic variables. Here, we examine these peculiar objects individually. Coordinates are  from 2MASS (J2000). Periods $P$, time-average $V$-band magnitude $V_{\rm CSS}$ and amplitude $\Delta V$ are from the Catalina catalog.\\
 
1) KG-03: ($\alpha,\delta$) = 00$^h$ 59$^m$ 58.93  $-$33\degree\, 28$'$ 35.1;
  $P$=196 days, $V_{\rm CSS} = 16.2$;  $\Delta V$=1.1 mag. This star is in the Sculptor dwarf galaxy (ALW 3 in Simbad) found by Azzopardi et al. (1997) [6] .\\

2) KG-05:  ($\alpha,\delta$) = 01$^h$ 01$^m$ 53.47 $-$65\degree\, 18$'$ 23.3;
$P$=162 days; $V_{\rm CSS} = 15.70$; $\Delta V$=0.5 mag.  This star is a known C-type star in the halo of the Magellanic Clouds, namely [KID97] 079-024 (Kunkel et al. 1997) [7]\\

3) KG-21: ($\alpha,\delta$) = 04$^h$ 56$^m$ 31.15 $-$31\degree\, 29$'$ 32.7;
  $P$=147 days; $V_{\rm CSS} = 14.26$ , $\Delta V$=1.4 mag. This star was already known (Mauron et al. 2018)[8]. It is not yet present in the Simbad database (CDS). Since the Mauron et al. paper was published recently, it is also not in the HG list.\\

4) KG-62: ($\alpha,\delta$) = 21$^h$ 26$^m$ 50.92 $-$37\degree\, 03$'$ 50.0;
 $P$=225 days;  $V_{\rm CSS} = 14.29$ , $\Delta V$=1.5 mag. This is a new carbon star discovery. It is not in the  listing of HG, nor in CDS.\\

5) KG-67: ($\alpha,\delta$) = 21$^h$ 43$^m$ 41.14 $-$34\degree\, 14$'$ 31.1;  $P$=158 days; ; $V_{\rm CSS} = 14.78$ , $\Delta V$=0.47 mag. This is a new carbon star discovery, neither in the listing of HG, nor in CDS.\\

We note that KG-62 and KG-67 are separated by a few degrees, and that there is no dwarf galaxy in this field. Comparison of their positions with the main locus (displayed in Fig.~1) of the tidal arms of the Sagittarius dwarf galaxy strongly suggests that these two new C stars originate in the Sgr dwarf. Confirmation with radial velocities is necessary.

\subsection{Peculiar M-type stars}

We discuss here three cases of M-type stars, before treating in the next section the subject of light contamination by bright, angularly close objects. The first star KG-08 is member of Fornax, and is useful for obtaining a typical absolute magnitude. The others are KG-63 and KG-83.

KG-08: When cross-matching the Catalina coordinates with  the point-source catalog of  2MASS (Strutskie et al. 2006) [9], the only 2MASS object within 4 arcseconds is at ($\alpha,\delta$) = 02$^h$ 39$^m$~15.33 $-$34\degree\, 15$'$ 08.3, and this identification is secure. This star was named F25006 by Whitelock et al. (2009) [10], when they performed a near-infrared $JHK$ monitoring of stars in Fornax. They find F25006 variable, but no period is given. The 2MASS single-epoch catalog provides $K_{\rm s} = 13.62$ $\pm 0.049$ while Whitelock et al. give a time-averaged $K_s = 13.74 \pm 0.25$, in good agreement. Catalina finds it to be periodic with $P=196$ days, and its phased light curve is of satisfactory quality. The 2MASS $J-K_s$ is 1.18, which is consistent with our M-type classification. Because the distance of Fornax is known (147 kpc), this star allows us to derive an absolute magnitude for this period: one finds $M_{\rm K} = -7.1$.

KG-63: The J2000 coordinates in the Catalina catalog are $\alpha = 322.63795$, $\delta = -53.96363$ (in degrees). An interrogation of the USNO-B1 catalog (Monet et al. 2003) [12] shows that within 10 arcsec, there is only one matching object, located at 0.39 arcsec from the Catalina position, and at $(\alpha, \delta)$ = 21$^h$ 30$'$ 33.14, $-$53\degree\, 57$'$ 48.8 (J2000). Examination of the POSS Digital Sky Survey images confirms this identification (see Figure 4). The $R$-band magnitude of this USNO-object is 18.9. The DR2 Catalina light-curve data are good and a value $P=223$ days is given by the DR1 (Drake et al.) catalog. The mean $V$-band Catalina magnitude is $V$\,=\,17.5, and the amplitude is 1.5 mag. If this star is in fact a semi-regular SRa variable, it is at a considerable distance. This star has a  faint counterpart on the 2MASS survey images, and the 2MASS point-source catalog gives at the above coordinates an object with $K = 15.60 \pm 0.2$, clearly seen in the 2MASS images. Adopting this $K$ and the $K$-band absolute magnitude of KG-08 (since periods are nearly equal), we obtain a distance of 350~kpc. This is astonishingly large and requires further study. 

By interrogating the GCVS (Samus et al. 2017) [11] for which we did not find any error in the coordinates, it is very intriguing that there exists a bright Mira located 51 arcsec North-East from KG-63 and with the same period: the Mira is X Ind. Its $V$-magnitude is between 8 and 13 as given in the GCVS, with a period  of 226 days, with $\Delta V = $ 5 mag. This Mira is clearly seen on the POSS plates (see Figure 3), but it is not present in the Catalina catalog because of saturation.

KG-83: In Catalina, the dataset is of satisfactory quality, with $V_{\rm CSS} = 16.8$, period of 255 days,  and amplitude 0.9 mag. This star is in 2MASS at $(\alpha, \delta)$ = 22$^h$ 43$'$ 11.06, $-$41\degree\, 31$'$ 12.9 (J2000), 
$K_s =15.25 $, $J-K_s = 0.71 \pm 0.16$. The same position is found in  USNO-B1.0 , where a $R$-band magnitude $R_1 =18.3$ is given. There is no doubt of the identification of KG-83, which would be at a considerable distance if a true AGB star. Again, in the GCVS, we find a Mira at 46$''$ from this object. This Mira is DS Gru, period 259 days, amplitude from 9.8 to 15 in the V-band.

These two cases, where the faint periodic Catalina star is found in the vicinity (at $\sim$ 1 arcmin) of a bright GCVS Mira with both having the
same period, suggest that variability of these faint objects is an artefact. We think that the variability of the faint Catalina star is due to the bright nearby Mira which saturates the Catalina detector and has a strong polluting halo. In other words, the faint Catalina star is a false long-period variable, although it does correspond to a star catalogued in USNO. In order to know more about this effect, we investigate below its frequency and characteristics in the whole (north and south) Catalina LPV catalog.

\section{Artefacts}

We consider on one hand the northern and southern Catalina long-period variables. On the other hand, we consider the GCVS, and we search for GCVS neigbhours (within 200 arcsec) of Catalina LPVs. There are of course objects which are in both catalog and at the same coordinates, within $\sim 3''$ which is the precision of coordinates from Catalina. GCVS coordinates are excellent, in general better than 1 arcsec, when a comparison is made with USNO or 2MASS or Gaia coordinates. We found as many as 56 Catalina cases where a bright LPV is in the 200$''$ field. All these but one are Miras. They lie at $\sim 1$ arcmin of the Catalina object, and with near-identical period. 

In Fig. 4, we show the histogram of angular separation (left panel). It is seen that it peaks at $\sim$~50$''$, and very few pairs have separations between $100''$ and $200''$. In the middle panel, we show the $\alpha$, $\delta$ offsets, and again the odd angular distribution suggests an instrumental effect. The right panel displays the histogram of period differences. Most of them are $< 5$ days, a value close to the typical error in the period determination. These panels show that it is a relatively frequent occurrence to find such false long-period variables. Several cases occur where 2 or 3 false LPVs surround the same bright Mira. For example, there are 4 false LPVs around AQ Cen, with separations of 135, 107, 92 and 59 arcsec.

\section{Discussion and conclusion}

Our original goal was to perform spectral classification of Catalina southern LPVs.
The method is to scrutinize objective-prism spectra offered by the Hamburg-ESO digital survey. Because this survey is based on blue plates and is limited to high galactic latitude ($|b| > 30$\degree), only a fraction of Catalina objects can be studied. Nevertheless, we could surely classify 83 cases out of 135 Catalina LPVs located in this region of the sky. We found 5 carbon stars. Two are new discoveries. The rest are M-type. Figure 1 shows that they are not concentrated along the location of the Sagittarius tidal arms, but instead are found in the $\alpha = 300$ to 360\degree~region of RA. The origin of these M-type LPV stars is therefore not explained with the available information. We cannot distinguish a thick disk or a halo membership. Further attention should also be paid to the 52 unclassified stars. 

While searching for very distant Catalina LPV stars, as faint as $K = 15$, which corresponds to 250 kpc if $M_{\rm K} \sim -7$, we found a large number of catalogued LPVs relatively close ($\sim$\,1\,arcmin) to bright Miras of the GCVS, and with the same period. The best explanation that we propose is that the variability of faint objects is due to contaminating light.  


In conclusion, we find that the overwhelming majority of LPVs in the southern hemisphere and at $|b| > 30$\degree\, are M-type stars. We also recommend researchers to take into account the polluting effect described here when seeking to identify very distant LPVs, and also to verify that no large amplitude bright LPVs are located within $\sim 3$ arcmin.

\section{Acknowledgments} The authors thank our referee, Prof. N.N.\,Samus, for many remarks that greatly improved the manuscript. We also thank Dr D.\,Engels for help in using the Hamburg database, and Dr. G.\,Rudnitskij for writing  the russian abstract. K.S.G. thanks CNRS, LATMOS, University of Versailles Saint Quentin en Yvelines, and LAM (Marseille) for supporting this study. This research has made use of the 2MASS and USNO catalogs, the SIMBAD and Vizier databases operated at CDS (Strasbourg) and the superb Catalina sky survey database operated at the California Institute of Technology, and funded by the US National Science Foundation and NASA.


\section{References}

\small

1. A.P. Huxor, E.K. Grebel, 2015, MNRAS, 453, 2653\\

2. A.J. Drake, S.G. Djorgovski, M. Catelan, et al. 2017,  MNRAS, 469, 3688\\

3. D.  Engels, H.-J. Hagen, N. Christlieb, D. Reimers, F.-J. Zickgraf, 2001 in ASP Conf. Ser. 232, The New Era of Wide Field Astronomy, ed. R.G. Clowes, A.J. Adamson, \& G.E. Bromage (San Francisco, CA: ASP), 326\\

4. N. Christlieb, D. Reimers, L. Wisotzki, 2004, The Messenger, vol. 117, 40\\

5.  A.J. Drake, M.J. Graham, S.D. Djorgovski, et al. 2014, Astrophys. J. Suppl. Series, 213, 9\\

6. M. Azzopardi, J. Lequeux, B.E. Westerlund, 1986, A\&A, 161, 232\\

7. W.E. Kunkel, M.J. Irwin, S. Demers, 1997, Astron. Astrophys. Suppl. Series, 122, 463\\

8. N. Mauron, K.S. Gigoyan, G.R. Kostandyan, 2018, Astrophysics, 61, 101\\

9. M. Strutskie, R.M. Cutri, R. Steining, et al. 2006, Astron. J., 131, 1163\\

10. P.A. Whitelock, J.W. Menzies, M.W. Feast, et al. 2009, MNRAS, 394, 795\\

11. N.N. Samus, E.V. Kazarovets, O.V. Durlevich, N.N. Kireeva, E.N. Pastukhova, 2017, Astronomy Reports, 61, 80\\

12. D.G. Monet, S.E. Levine, B. Casian, et al. 2003, Astron. J., 125, 984\\

\hspace{1cm}


\begin{figure}[!ht]
\begin{center}
\includegraphics*[width=8cm,angle=-90]{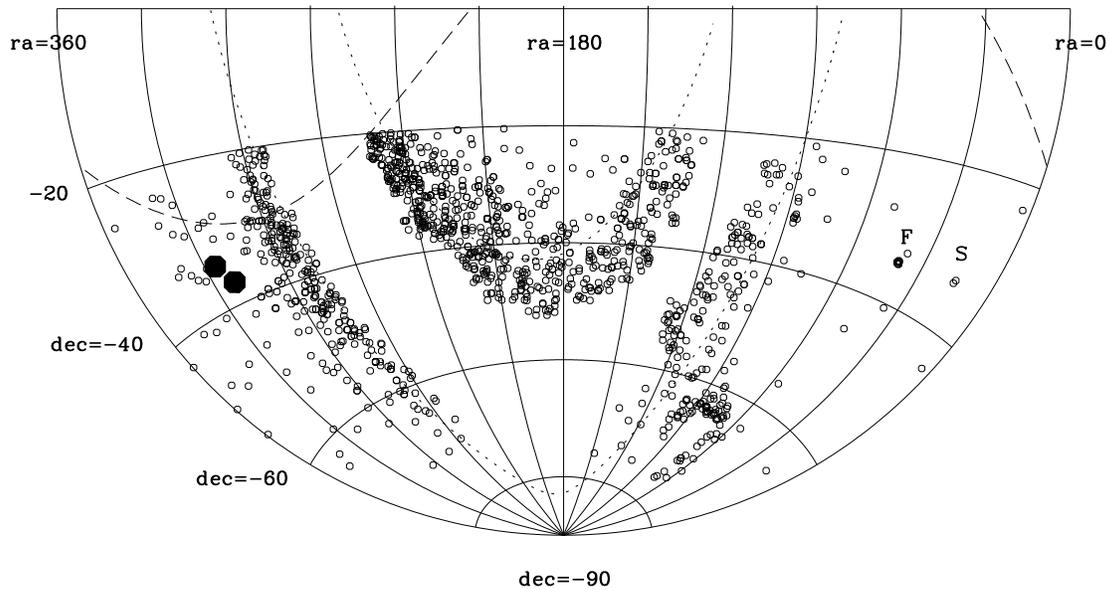}
\caption{Equatorial map R.A.-Dec. of 1286 Catalina long-period variables (circles). 
Dotted lines indicate  $b = +20$ or $-20$\degree\,. Long dashed lines show the average
location of the Sgr tidal arms. To the right, F and S mark the position of the Fornax and Sculptor dwarf galaxies, respectively. At lower right, the clump is due to the Large Magellanic Cloud. This plot shows that a considerable number of the Catalina LPVs are located close to the Milky Way, but that this experiment probes up to 150 kpc away, the distance of Fornax. The 2 new carbon stars are indicated by large filled hexagons.}
\end{center}
\end{figure}

\begin{figure}[!ht]
\begin{center}
\includegraphics*[width=16cm,angle=00]{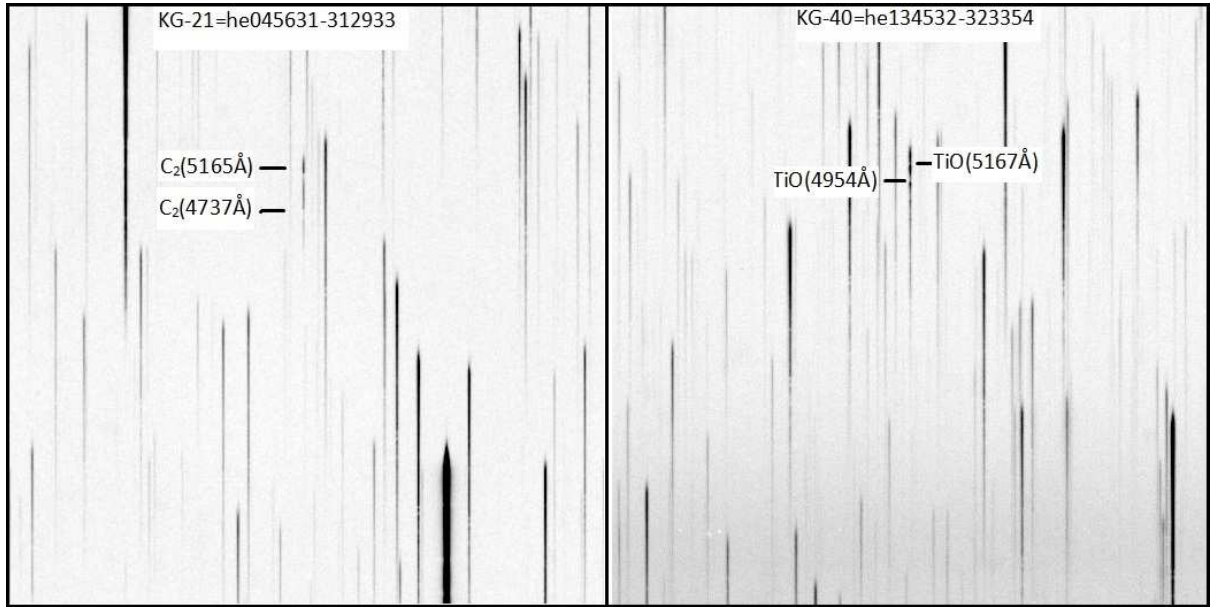}
\caption{HES low-resolution spectra in the range 3200-5300 \AA\, for the stars KG-21 (a C-type star, left panel), and for KG-40 (an M-type star, right panel). The molecular bands of the C$_2$  at 4737, 5165 \AA, and those of TiO  at 4954, 5167 \AA\, are indicated. Both fields are 15$'$ $\times$ 15$'$.}
\end{center}
\end{figure}

\begin{figure*}[!ht]
\begin{center}
\includegraphics*[width=8cm,angle=0, origin=c]{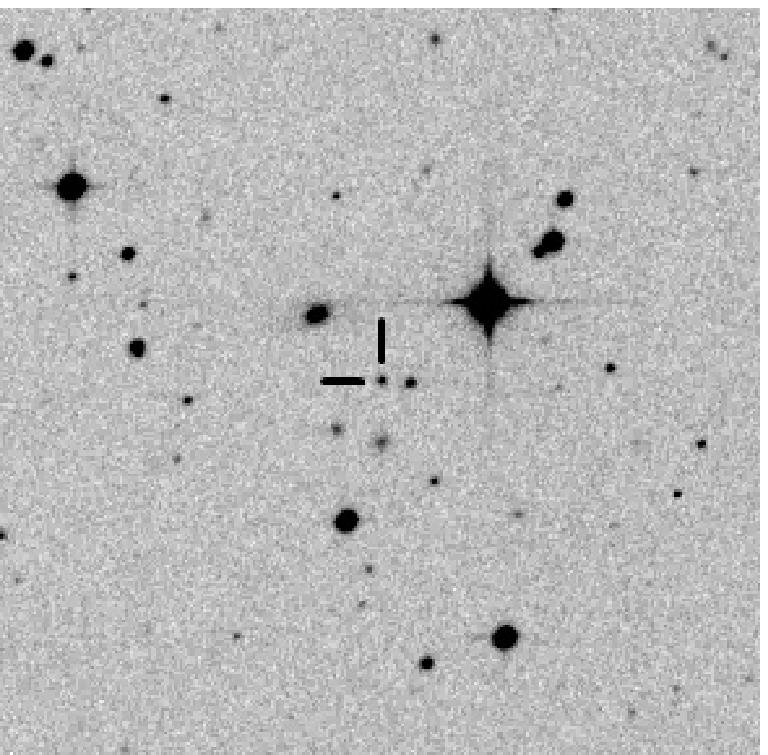}
\includegraphics*[width=8cm,angle=0,origin=c]{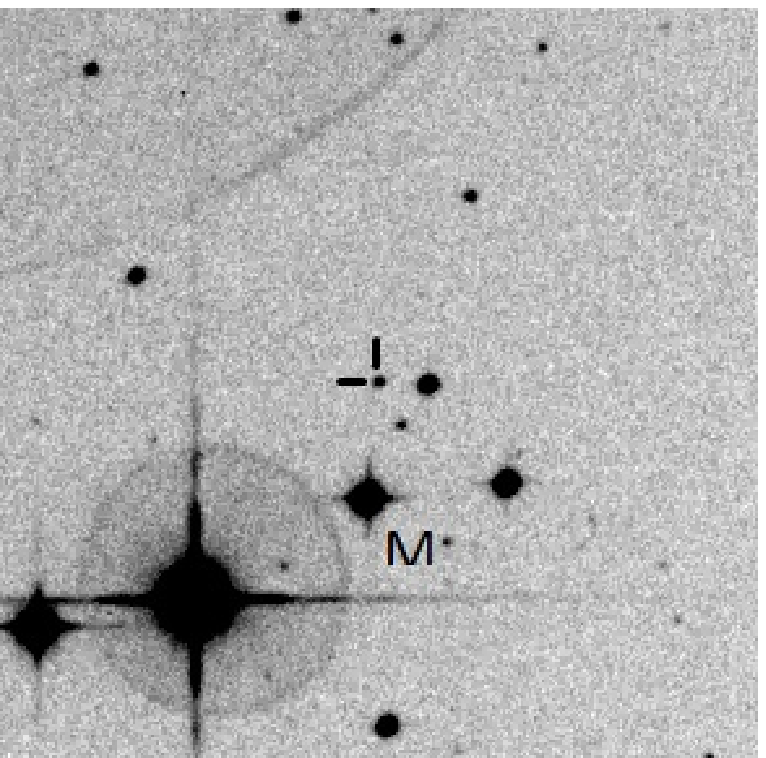}
\caption{Finding charts with size 300 arcsec, North is up, East to the left, from red Palomar Sky Survey digitized images. On the left panel, KG-63 is indicated by ticks and the Mira X Ind  is the brightest star in the field (angular separation 51$''$). One the right panel,  KG-83 is indicated by ticks and the Mira DS Gru is the third brighter star in the field and is indicated by the letter M (angular separation 46$''$)}.
\end{center}
\end{figure*}

\begin{figure*}[!ht]
\begin{center}
\includegraphics*[width=5cm,angle=-90, origin=c]{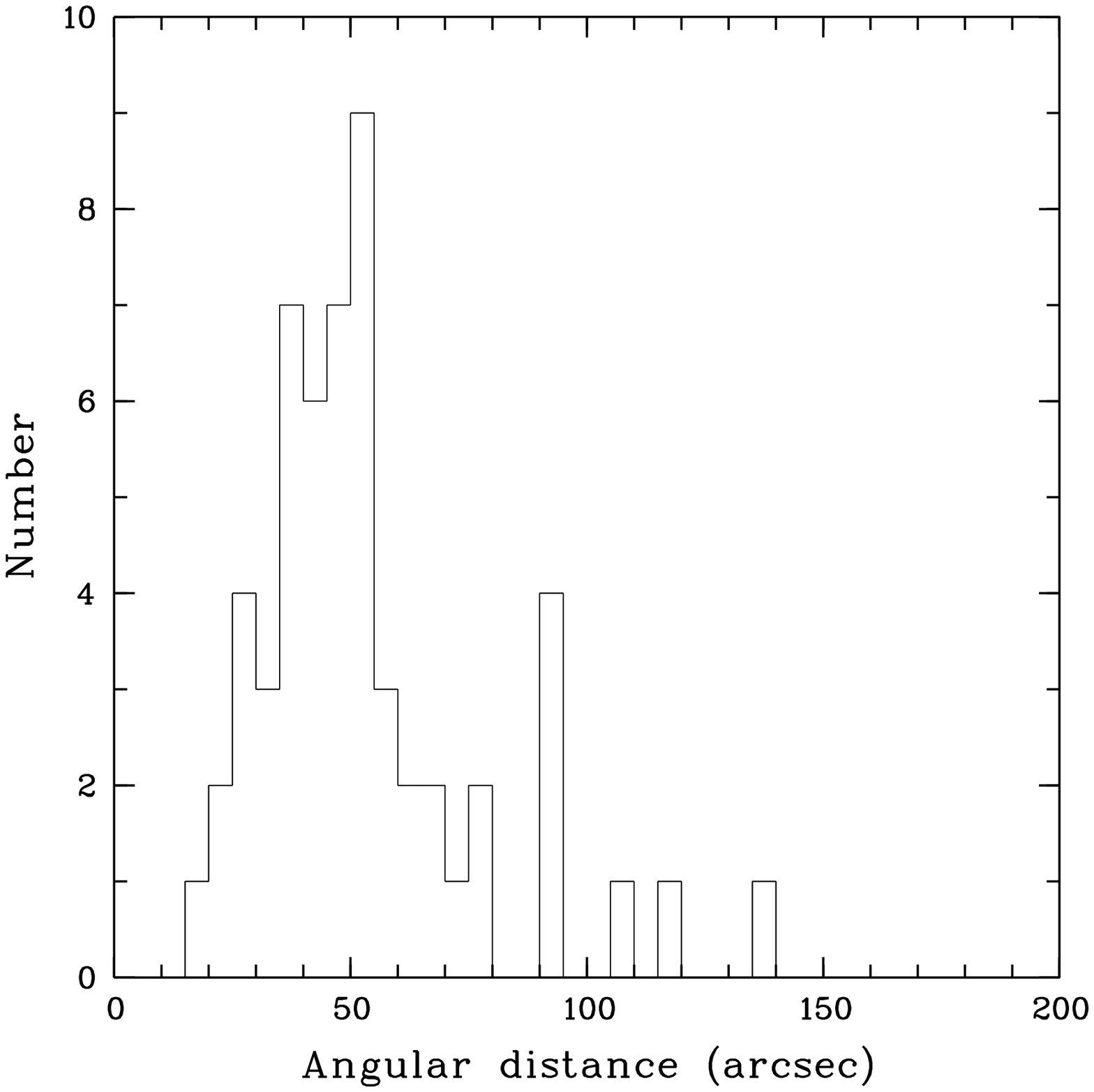}
\includegraphics*[width=5cm,angle=-90,origin=c]{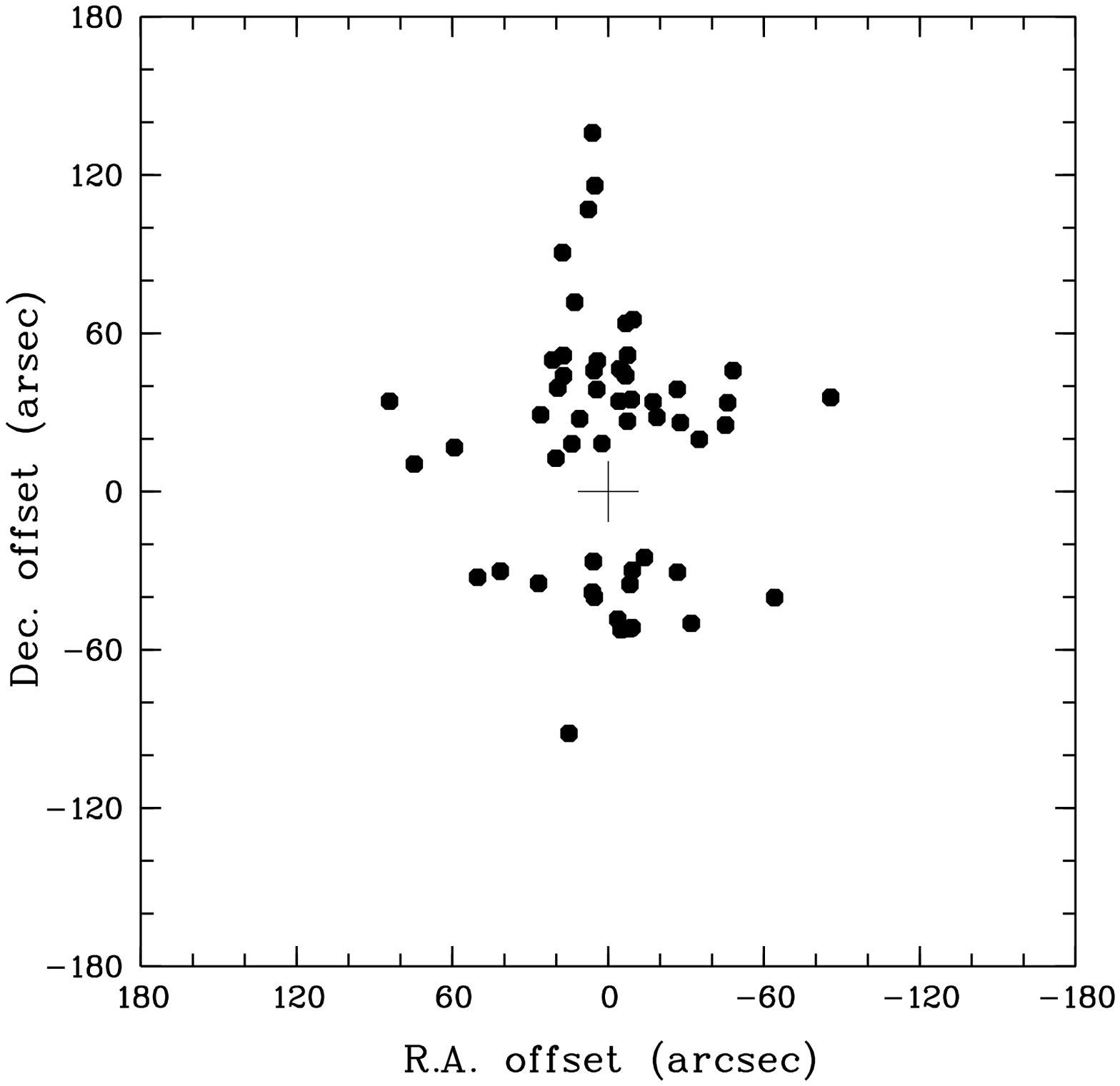}
\includegraphics*[width=5cm,angle=-90, origin=c]{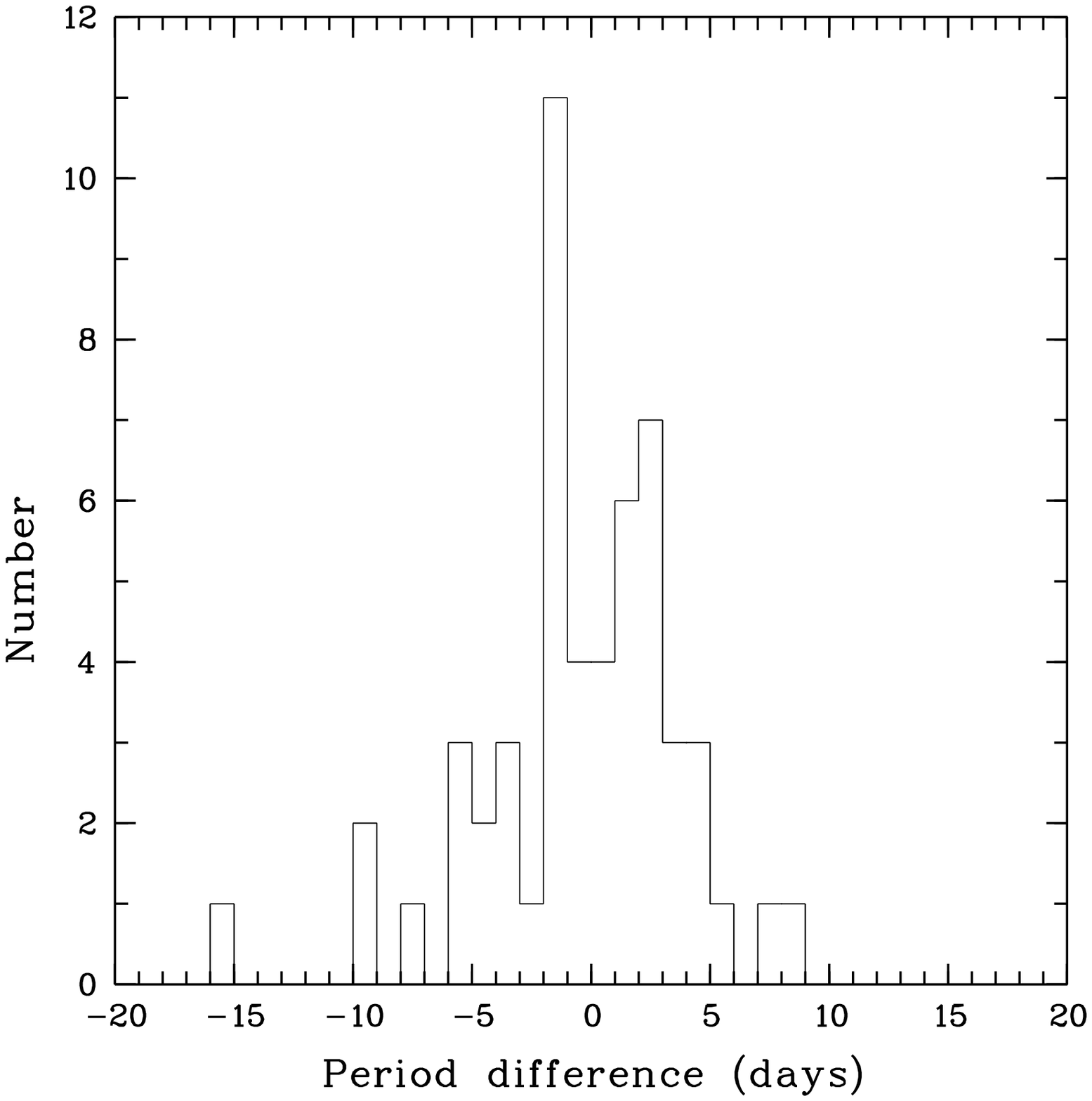}
\caption{Properties of the faint ``false" Catalina LPVS. {\it Left panel}: Histogram of the angular separation  between the faint Catalina Mira and the bright GCVS Mira. {\it Middle panel}: Positional offsets (in $\alpha$, $\delta$) of the faint LPVs with respect to the bright GCVS Mira (indicated by a cross). {\it Right panel}: Histogram of the difference between the period of the bright Mira   and the period of the faint Catalina object. (see text for more details) }
\end{center}
\end{figure*}


\begin{figure*}[!ht]
\begin{center}
\includegraphics*[width=18cm,angle=0, origin=c]{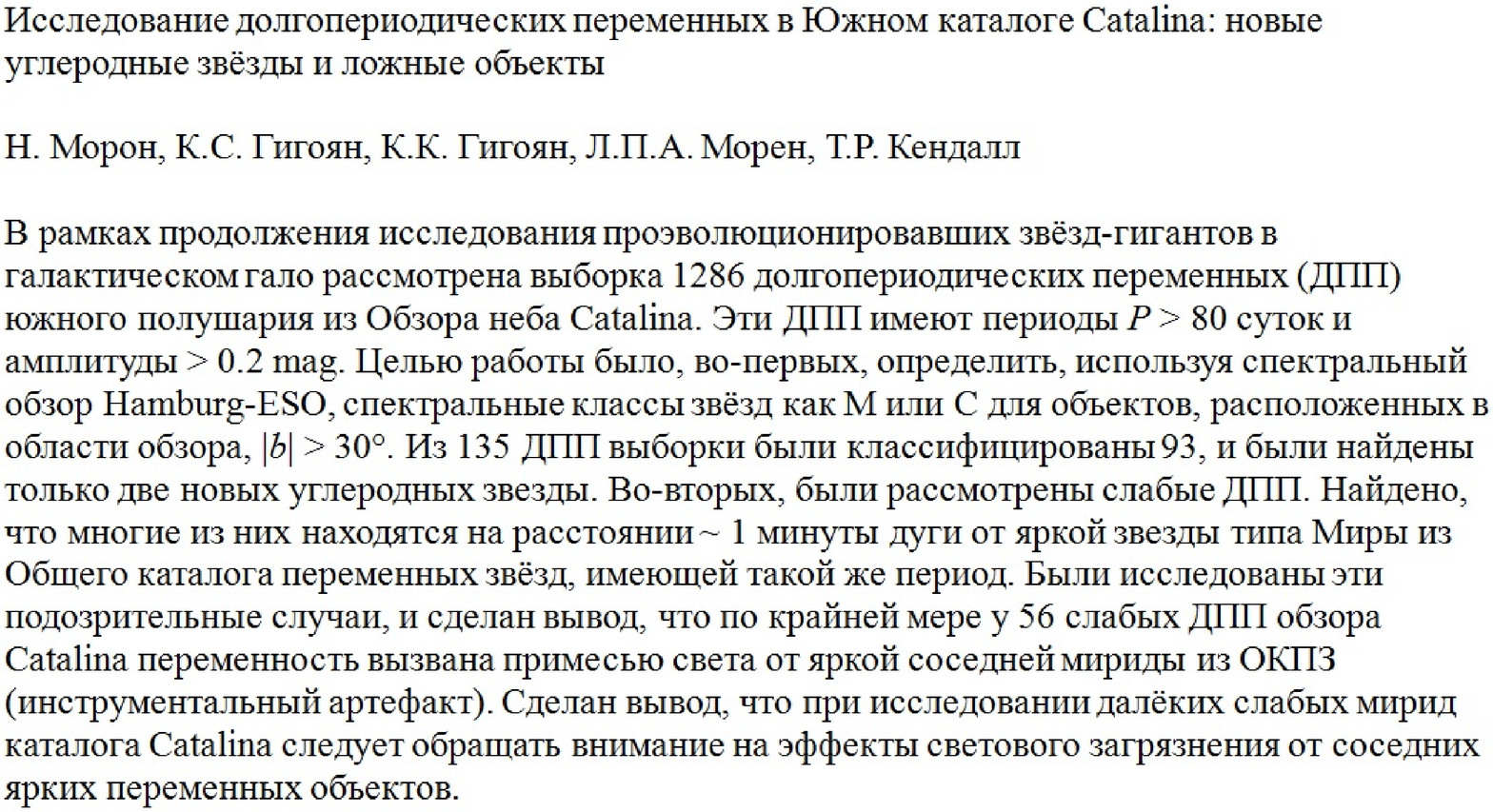}
\caption{Abstract in russian}
\end{center}
\end{figure*}

\end{document}